# Micro/nanoliter droplet extraction by controlling acoustic vortex with miniwatt

Han Zhang[1,2,*)], Jun Yang[1,2], Yun Zhou[3,4], Jianfeng Zheng[4], Yong Cheng[5], Bichao Bai[4], Guoxin Zhang[6], Yisheng Lv[7]

[1]CAS Key Laboratory of Noise and Vibration, Institute of Acoustics, Chinese Academy of Sciences, Beijing 100190, China.

[2]University of Chinese Academy of Sciences, Beijing 100049, China.

[3]The National Center for Nanoscience and Technology, Beijing , 100190, China.

[4]School of Mechanical Engineering, Changzhou University, Changzhou 213164, China.

[5]Institute of Mechanics, School of Civil Engineering & Mechanics, Yanshan University, Qinhuangdao 066004, PR China.

[6]Al-ion Battery Research Center, Department of Electrical Engineering and Automation, Shandong University of Science and Technology, Qingdao 266580, China.

[7]State Key Laboratory of Management and Control for Complex Systems, Institute of Automation, Chinese Academy of Sciences, Beijing 100190, China.

(*Corresponding author: zhanghan@mail.ioa.ac.cn)

## ABSTRACT

Micro/nanoliter droplet is capable of achieving versatile applications with tiny volume and substantial surface energy, which is a big plus over bulk liquid. Yet, the contradiction of elaborate manipulation and enough power is still a challenge. Here, we unleash the potential of our miniwatt aspirators pumping up liquid and creating droplets with the help of acoustic vortex beams, inspired by the power mechanism that spirals are significant for most mollusks that live in water. These droplet aspirators produce very large interface deformations by small radiation pressures with orbit angular momentum from spiral-electrode transducers. The precisely contactless manipulation of physical, chemical and biological objects at micrometric down to nanometric scales, promises tremendous development in fields as diverse as microrobotics, nanoreactors, or nanoassemblies.

*Keywords:*

Micro/nanoliter droplet

## 1. Introduction

Generation and manipulation of droplets have contributed in many outstanding achievements in physics, energy, biology, chemistry and material science over the latest few years [1-13]. During these successful experiences, droplet is prominently capable of achieving versatile applications with tiny volume and substantial surface energy in comparison with bulk liquid. In this regard, it is of great value actuated by droplet manipulation in the fields of micro-fluidic, liquid transportation, droplet-based electric generator, droplet-based printing, single droplet reactor, multifunctional sensors, etc. Paradoxically, almost without exception to the equipment in micromanipulation, the more precise they are, the more expensive their cost needs. What is even more rigorous is that their propelling powers have been inevitably restricted due to their dimensions. Yet, micro/nanoliter droplet is still a harsh challenge [5] in the process of preparation and manipulation, particularly precluded from separability and selectivity at the single droplet level [12-19]. Wave control totally overcame the definition of macroscopic dimensions, which has been a preeminent tool for micromanipulation dexterously enough at the wavelength level, through contactless forces and energy from wave motion. Among the pioneers, Lord Rayleigh in 1902 first put forward the description of radiation pressure when wave propagating in acoustics, optics or electromagnetism [20]. Then Chu employed the resonance radiation pressure of a set of six-laser beams to realize the targeted atom captured in three-dimensional viscous confinement [21]. In the atomic traps, the kinetic energy of the cooled atoms is reduced to a suspended state since exchanging momentum between photons and atoms. Arthur went a step further to enhance the gradient radiation pressure through a highly focused beam to demonstrate a single-beam gradient force trap of dielectric particles for the first reported time and throughout the concept of negative light pressure due to the gradient force was verified experimentally [22]. Not only extremely tiny objects and ultra-fast processing but also much more advanced equipment in the span-new fields for broad industrial and medical applications resort to the invention of optical tweezers. Up till now, acoustic tweezers have emerged as a prominent technique for controlling cell–cell interactions without deleterious heating while providing more noninvasive, more biocompatible, higher penetrability in organic tissues than previous techniques [23-26]. The most important thing is that they verified the contactless and damageless micromanipulation in wavelength level of three-dimensional acoustic radiation force [27-29]. So, acoustic wave control supplies us a potential tool [30-32] to breakout the harsh challenge in the preparation and manipulation process of micro/nanoliter droplets, which is mentioned above.

Here, we harness the mechanism to explore a new kind of miniwatt aspirators to pump up liquid and create droplet(s) via acoustic vortex beams, inspired by the power mechanism that spirals are significant for most mollusks that live in water [33]. It is the nature hidden in their power mechanism with the minimum energy consumption [34-36] that a tiny change in polar angle can result in a huge change in polar radius over a certain value of the spiral coefficient. The resistance from the flow is transformed into the propulsion by the pressure difference also due to their spiral structures [37]. Acoustic vortex beams possess both the advantages of the spiral structures and the advantages of the precise wave control, which perform an astonishing variety of applications in many dynamic manipulation [7-11]. However, they have not been applied in the droplet manipulation of steady extracting yet. Based on the effects of acoustic radiation pressure on two-phase liquid interfaces [38-40] and orbital angular momentum (OAM) transferring from the spiral waves spinning around a phase singularity [41-43], we achieve this kind of dynamic manipulation with our smart invention of miniwatt droplet aspirators illustrated in Fig. 1.

By the spiral dynamic mechanism of vortex beams, we succeed in resolving the contradiction of precision and power in wave control. Hence, we present 5 parts to address our miniwatt droplet aspirators in this paper. (I) We set up the model of our invention in cylindrical coordination as Fig. 2(a) shown and give out the theoretical analyses of the acoustic radiation force and the acoustic pressure field of spiral-electrode transducer. (II) We obtained the deformations of the two-phase-liquid interface and then the droplets with wavelength-scale in the experiments designed as Fig. 2(b). (III) We verified our miniwatt power costs in the comparison of transducers with non-spiral and spiral electrodes (Fig. 3). (IV) We employed Bernonlli theorem to demonstrate the contribution of OAM transfer to saving power cost, which is benefits to the rotation velocity of liquid in the upper layer. The pressure of the upper fluid against the lower fluid reduced by the existence of OAM transfer is illustrated in Fig. 2(b). (V) Finally, we analyze the force of the slant liquid column produced on the liquid interface when the transducer is slant. The minimum power of our aspirators is optimized by the slant vortex beam at 45° which realized our further optimization of the power costs.

## 2. Theory models and design methods

In this paper, the control of liquid level and droplets mainly depends on the wave control of acoustic waves. The mechanism of micro/nanoliter droplet aspirators based on vortex beams is depicted in Fig. 1. The bottom red medium is carbon tetrachloride ($CCL_4$) and the upper blue region represents the water. Spiral piezoelectric transducer is adopted to produce vortex acoustic field. Therefore, the $CCL_4$ droplet could be manipulated by the acoustic radiation force.

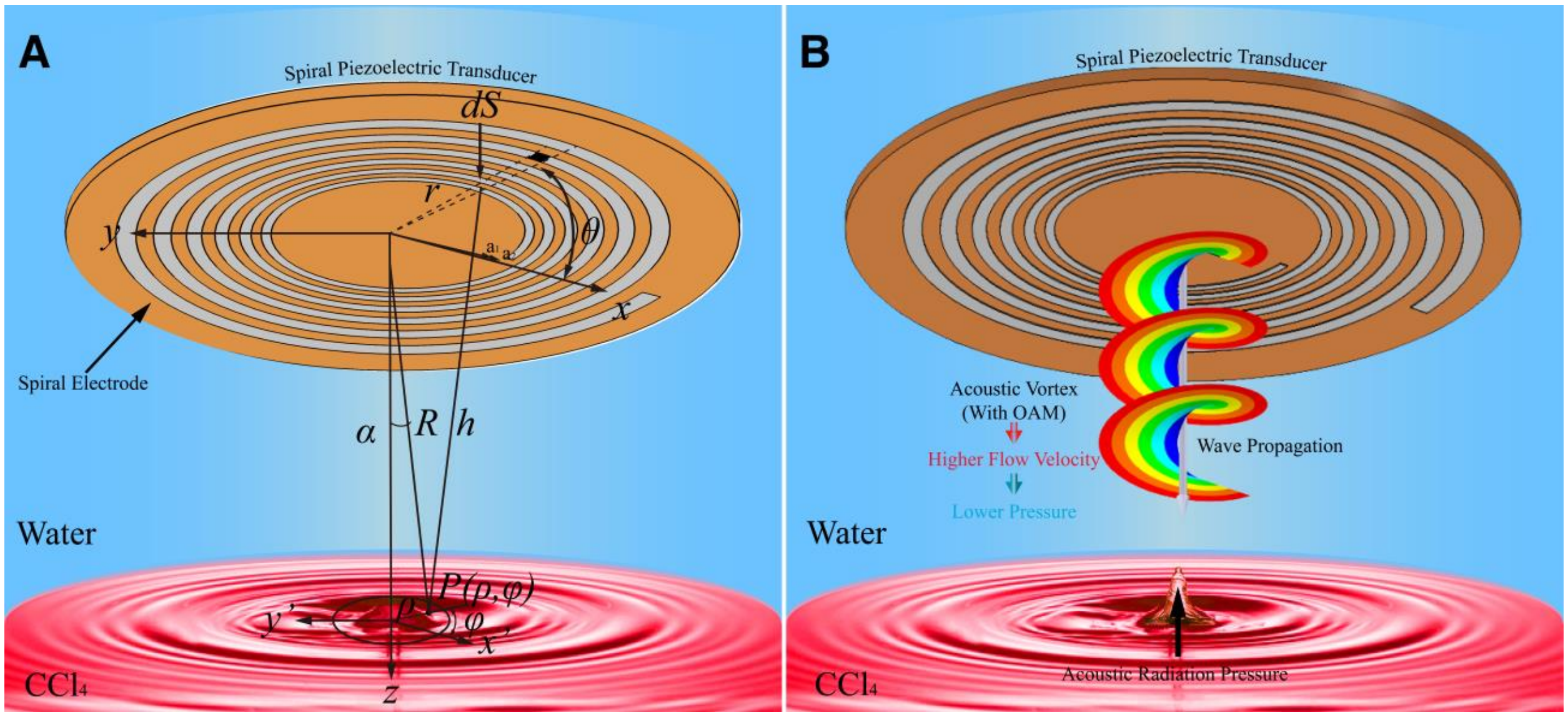

FIG. 1. Schematic diagram of the miniwatt spiral aspirators (a) cylindrical coordinate; (b) vortex acoustic field

According to the design of the spiral electrode on the surface of the transducer, the acoustic pressure field produced by the single spiral electrode transducer under the cylindrical coordinate can be expressed as,

$$p(\rho,\varphi,z)=v_a\rho_0 c_0\int_0^{\theta_{\max}}\int_{a_1e^{b\theta}}^{a_2e^{b\theta}}\frac{i\omega\rho_0}{2h}e^{i(\omega t-kh)}rdrd\theta \tag{1}$$

where $p$ represents the acoustic pressure at point $P$ in the acoustic field, $\rho$ represents the radial distance, $\varphi$ represents the azimuth, $z$ represents the axial distance, $v_a$ represents amplitude, $\rho_0$ and $c_0$ represent the density and acoustic velocity of water, $h\left(=\sqrt{\rho^2+r^2+z^2-2rR\cos\varphi\sin\alpha}\right)$ represents the distance of $dS$ on spiral electrode to point $P$, $R\left(=\sqrt{\rho^2+z^2}\right)$ represents the distance from the center of transducer surface to point $P$ and $\alpha$ represents the angle between the axis and the connection line of point $P$ and the center of transducer surface. We integrated the whole area of the spiral electrode, and the integral terms $r$ and $\theta$ in the integral equation represent the radius and azimuth of the d$S$ on the spiral electrode, respectively.

The Brillouin's radiation stress tensor $S_T$ under the cylindrical coordinates can be expressed as,

$$\mathbf{S}_T = \left(\langle K\rangle - \langle V\rangle\right)\mathbf{I} - \rho_0\langle\mathbf{uu}\rangle \tag{2}$$

where $\langle \bullet \rangle$ represents the volume average of the corresponding quantities of the acoustic field in Fig. 1(a). $K$ and $V$ represent the kinetic energy and the potential energy, respectively. $\mathbf{I}$ is the unit tensor, and $\mathbf{u}$ is the particle vibration velocity at point P. According to the definition of the kinetic energy and the potential energy, the above expression can be calculated as,

$$\mathbf{S}_T = \left(\frac{\rho_0}{2}\langle\mathbf{u}^2\rangle - \frac{1}{2\rho_0 c_0^2}\langle p^2\rangle\right)\mathbf{I} - \rho_0\langle\mathbf{uu}\rangle \tag{3}$$

Then the radiation force resulting from the vortex acoustic field can be expressed as the integral of the Brillouin stress tensor,

$$F = \oint_A \mathbf{S}_T \cdot dA = -\oint_A \rho_0\langle\mathbf{uu}\rangle\cdot dA + \oint_A \left\{\frac{\rho_0}{2}\left\langle|\mathbf{u}|^2\right\rangle - \frac{1}{2\rho_0 c_0^2}\left\langle p^2\right\rangle\right\} dA \tag{4}$$

From $\mathbf{u} = \nabla\psi$ and $p=\rho_0\partial\psi/\partial t$, the particle vibration velocity $\mathbf{u}$ can be obtained by the acoustic pressure $p$, where $\psi$ represents the velocity potential. Therefore, the solution of the radiation force depends on the expression of acoustic pressure generated by the single spiral-electrode transducer. The upper and lower limits of $r$ are the outer radius and inner radius of the spiral electrode, respectively.

According to the study of the radiation force of the spiral Bessel beam by Marston [44], the axial radiation force $F_z$ can be expressed as $F_z=\pi\rho_0 k(I_3-I_1-I_2)$ (Supplementary I). $I_1$ and $I_2$ are the components obtained by the potential energy part in Eq. 3, respectively. $I_3$ is the component of the product of velocity of particle vibration, and it can also be regarded as the component of the product of velocity potential gradient. Since $I_3$ has a positive effect on the axial radiation force, that is, the larger the $I_3$, the greater the axial radiation force. The radiation force is negative when extracting the liquid column at the two-phase liquid interface, so the larger $I_3$ is, the smaller the negative radiation force is. $I_3$ is the component of the velocity potential gradient product. It is well known that the acoustic field of the acoustic vortex is a kind of spiral acoustic wave, and there is a certain angle in the acoustic wave direction. Compared with the ordinary plane wave moving in a single direction, it will have a larger velocity potential gradient, so its radiation force could be smaller.

## 3. Experimental demonstration of precise extraction of droplets

Our experiment was carried out in a transparent cylindrical glass container with a height of 30 cm and a diameter of 20 cm, as shown in Fig. 2(b). The two kinds of liquids used in the experiments are carbon tetrachloride ($CCl_4$, density $\rho_1$=1600 kg·m$^{-3}$, acoustic velocity $c_1$=926 m·s$^{-1}$) and water (density $\rho_2$=1000 kg·m$^{-3}$, acoustic velocity $c_2$=1500 m·s$^{-1}$). Because of $\rho_1 \cdot c_1 \approx \rho_2 \cdot c_2$, the difference of acoustic impedance between the two liquids is very small, and acoustic waves can pass through almost lossless. In terms of equipment, we use a signal generator to output low-power RF signals and then use a power amplifier with a maximum magnification of 50 dB to gain high-power RF signals that required for the transducer to work properly. The working power of the piezoelectric transducer is also controlled by the signal generator. Oscilloscope was connected to both ends of the transducers to better measure the real-time working voltage of the transducers during the experiments. Except the measure of voltage, we combined the virtual instrument board NI ELVIS □+ of NI Company and the digital multimeter on the computer to measure the high-frequency current in the experimental circuit in real time. Due to the rapid process of the deformation of interface and extraction of droplets, we used a high-speed camera to shoot experimental phenomenon and computer to record videos in order to better observe and record the test phenomenon. In the experiment, the single spiral-electrode transducer is fixed in the upper water by a fixture, its side with the spiral electrode faces down and is parallel to the interface of the two-phase liquid, so as to emit acoustic vortex to the interface [Fig. 2(b)].

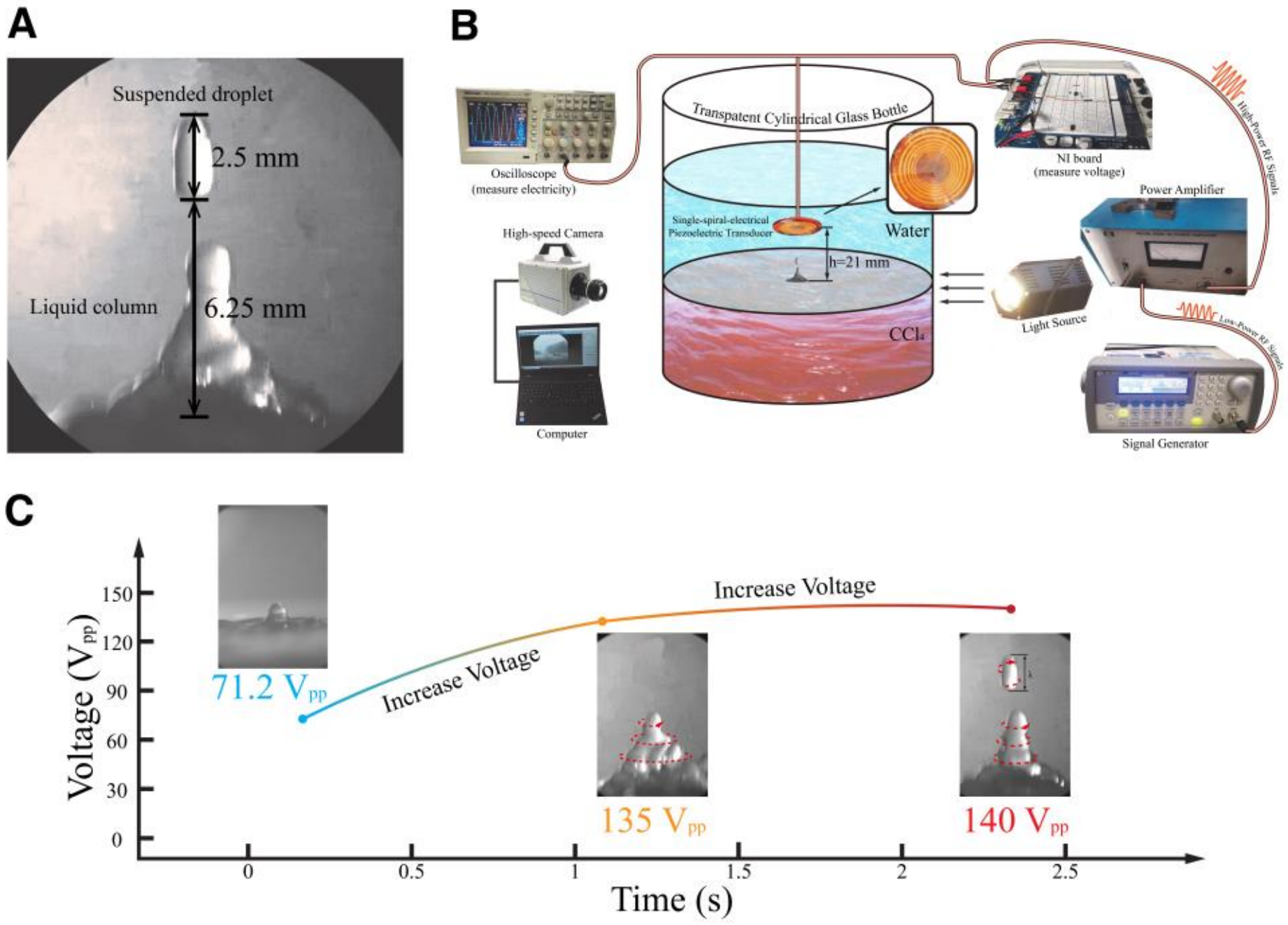

FIG. 2. The process of droplet appearance. (a) The liquid column extracted by spiral electrode transducer; (**b**) Schematic of experimental design and devices. (c) Process of extracting droplets by single spiral-electrode transducer.

According to the single spiral-electrode transducer used in our experiment, the inner radius of the spiral electrode is $a_1e^{b\theta}$, the outer radius is $a_2e^{b\theta}$, Where initial inner radius $a_1$ = 9 mm, initial outer radius $a_2$ = 9.65 mm and growth rate $b$ = 0.0025. The range of $\theta$ is 0 to 12 π, which means that the spiral electrode rotates 6 times totally.

Since the single vortex is the most basic and stable structure in all acoustic vortex, we used a single spiral-electrode transducer [Fig. S1] which can produce single vortex. In the experiment, we successfully realized the deformation of the interface of two-phase liquid at 608.3 kHz [Fig. 2(c)]. At this time, the distance between the lower face of the transducer and the liquid level is 21 mm, the working voltage is 71.2 $V_{pp}$, the current is 0.27 A, and the power is 6.8 W. After the single spiral-electrode transducer successfully aspirating the $CCl_4$ liquid column, we increased the voltage applied to the transducer, and the height of the liquid column increased as the voltage goes up. When the working voltage of the transducer increases to 140$V_{pp}$, the top of the liquid column becomes unstable and separated independent $CCl_4$ droplet. After maintaining the voltage for about 1 s, the single spiral-electrode transducer successfully captured the independent droplets detached from the liquid column and suspending them stably in the water layer, as shown in Fig 2(a). At this time, the working voltage of the transducer is 140$V_{pp}$, the current is 0.8A and the power is 39.6W. The droplets captured in the experiment are oval, the length of the major axis is 2.5mm and the length of the minor axis is 1.2mm. The size of droplet is close to the wavelength of acoustic waves in water (2.46mm), and the volume of droplet is about 1.9$mm^3$. We have effectively proved the feasibility of the droplet aspirator through experiments. The droplet aspirator based on acoustic vortex shows an ideal working state from promoting the deformation on the liquid interface at low voltage to accurately capturing and manipulating suspended droplets at high voltage. In order to further illustrate the unique advantages of our designed system, we combined with simulations to explain the significant advantages of our droplet aspirator based on acoustic vortex in liquid extraction and capture through the calculation of acoustic field and power.

In addition to the theoretical calculation, we also combined it with the COMSOL simulations to further research the acoustic field of the transducer. The working frequency of the transducer in simulation is 608.3kHz. According to the simulation results, we calculate the acoustic power at the surface of the transducer, which is 2.61 W. The power of 2.61 W we use by vortex beams is less than a power of 7.2 W used in [45] by focused beams.

According to Eq. 4, we can solve the radiation force produced by the single spiral-electrode transducer on the liquid interface combined with the simulation results. When the parameters in the simulation results are brought into the solution equation of the radiation force, the time average of the parameter product in the equation can be expressed as the product of the corresponding first complex number and the corresponding second conjugate complex number.

$$P_r = -\rho_0\left(\mathbf{uu}^*\right) + \frac{\rho_0}{2}\left(|\mathbf{u}||\mathbf{u}^*|\right) - \frac{1}{2\rho_0 c_0^2}\left(pp^*\right) \tag{5}$$

According to the Eq. (5), we can obtain the distribution diagram of $|P_r|$ at the interface of two-phase liquid. At the same time, we also add the simulation results of the ordinary circular electrode transducer without spiral-electrode, to compare the $|P_r|$ generated by the two transducers on the liquid interface [Fig. 3(a)]. Due to non-spiral transducer has a larger electrode area which lead to larger current at the same voltage, the working power of transducer will be larger. In order to maintain the same working power of two kinds of transducer during the comparison, we reduce the simulation voltage of the non-spiral transducer (Supplementary IV).

However, in the experiment, we found that at the same working power, the focused beam produced by the non-spiral transducer is not as effective as the single spiral transducer in extracting droplets. Therefore, it shows that the acoustic vortex does not only rely on the acoustic radiation force when extracting the droplet, and the biggest difference between the focused beam and the acoustic vortex is whether it carries angular momentum or not. Therefore, the angular momentum carried by the acoustic vortex plays an important role in the process of extracting droplets.

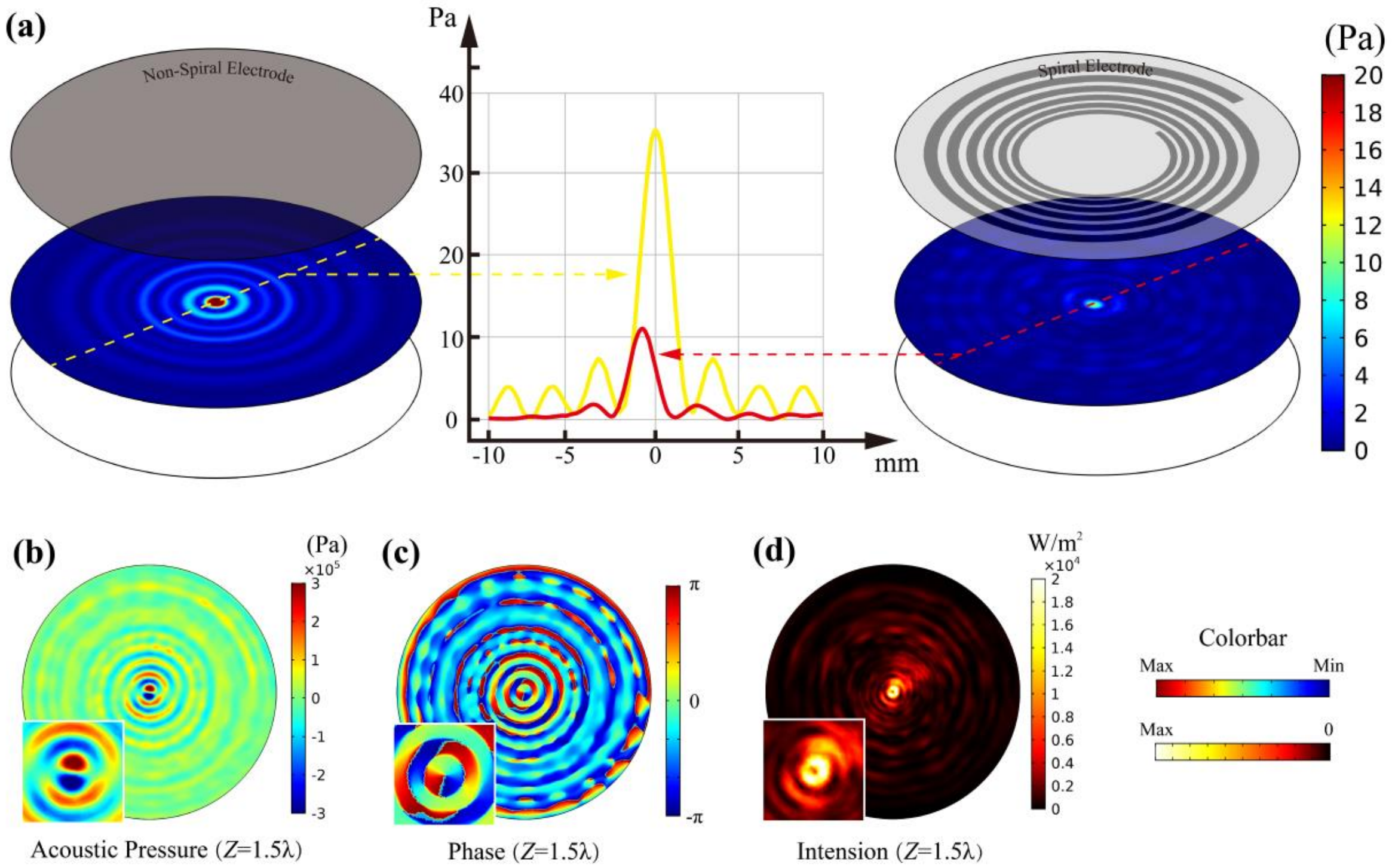

FIG. 3. Comparison of acoustic radiation forces and diagram of acoustic vortex beams (a) Comparison of acoustic radiation pressure at the interface between non-spiral electrode and single spiral electrode transducers; (b, c and d) The acoustic pressure, phase and intensity at 1.5 wavelengths from the surface of the spiral transducer with the cross-sectional views.

Actually, the vortex can be regarded as the winding and superposition of two acoustic beams with phase difference of π. We can better observe the special acoustic field of the vortex by observing the cross-section on the z-axis. Here, we intercept the plane distribution diagram of various acoustic field parameters at 1.5 wavelengths (3.7mm) from the transducer. From the cross-sectional view of the acoustic pressure, we can see that there are both positive and negative acoustic pressure in the center of the section, and the symbol of the two acoustic pressure are opposite [Fig. 3(b)], which represent the two acoustic beams with a phase difference of π. According to the cross-section of phase, we can see that the phase obviously makes a jump from π to -π [Fig. 3(c)], which is also the unique phase feature of the acoustic vortex. Finally, we also give the cross-sectional view of the acoustic field intensity [Fig. 3(d)], and the distribution diagram of acoustic pressure intensity is the time average of the acoustic intensity. It can be seen that a bright ring is formed in the center of the intensity field, and there is a small part of the zero intensity region at the center of the bright ring, which is due to the acoustic pressure offset at the center of the two beams with a difference of π in the vortex. Here, we introduce the distribution diagram of acoustic pressure, phase and intensity in the acoustic field of a single spiral-electrode transducer. It is explained from many aspects that the single spiral-electrode transducer can effectively produce the acoustic vortex with smaller radiation force.

Zhang et al analyzed and studied the transfer of OAM [46] and OAM carried by the acoustic vortex. Here, we further analyze the mechanism of lifting the liquid column of the spiral transducer based on the loss of angular momentum of the acoustic vortex and the Bernoulli principle. The Bernoulli equation can be expressed as,

$$P_h + \frac{1}{2}\rho v^2 + \rho g h = C \tag{6}$$

In Eq. (6), $P_h$ represents the pressure at a certain point, $v$ is the fluid velocity at the change point, $\rho$ is the fluid density, $h$ is the height at that point, and $C$ is a constant. This equation can be explained in a more easy-to-understand way: where the flow rate is fast, the pressure is low, and where the velocity is low, the pressure is high.

In our experiment of extracting liquid droplets, the acoustic radiation pressure produced by acoustic waves at the interface of two-phase liquid needs to overcome not only the surface tension of the liquid interface, but also the pressure given by the upper liquid. When the acoustic vortex produced by a single spiral-electrode transducer propagates in the upper water, the angular momentum carried by the acoustic vortex will be transferred to the water supply, thus speeding up the flow speed of the water. Then, according to Bernoulli's law, when the velocity of water increases, the pressure of the upper water on the interface of the two-phase liquid will decrease, so under the action of acoustic radiation pressure, the liquid interface is more likely to deform and form a protruding liquid column. Therefore, theacoustic vortex produced by the transducer will affect the velocity and pressure of the upper liquid through the exchange of angular momentum. The magnitude of angular momentum can be calculated according to Eq. (7).

$$\left\langle \dot{L}_Z \right\rangle = \frac{m \cdot P}{\omega_B} \tag{7}$$

where $m$ represents the topological charge of theacoustic vortex, $P$ represents the acoustic power, and $\omega_B$ represents the angular frequency of the acoustic beam. Therefore, there is no angular momentum in the transducer without spiral, that is, the acoustic beam produced by m ≠ 0, and only the acoustic vortex with m ≠ 0 carries angular momentum.

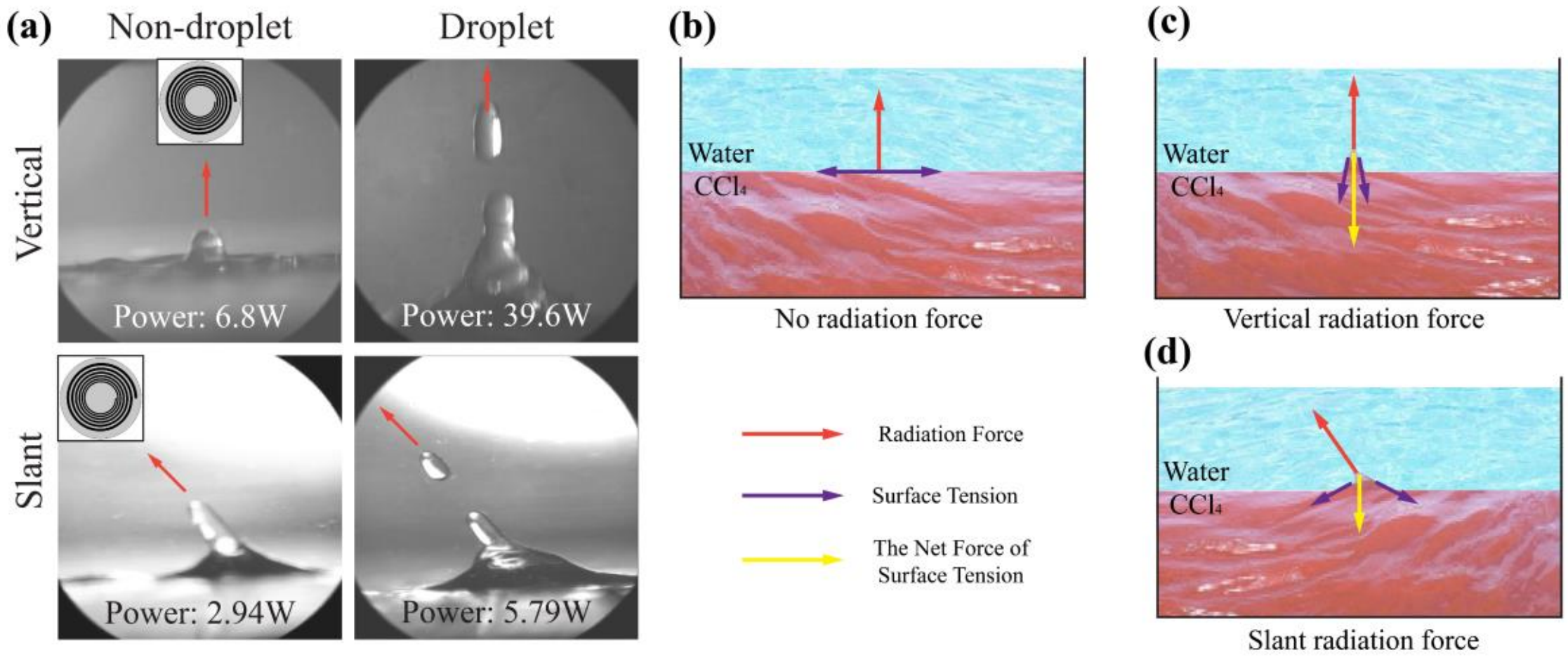

FIG. 4. The result when single spiral-electrode transducer is slant. (a) Power comparison of vertical and slant transducers in extraction column and droplet extraction; (b, c and d) Schematic diagram of force analysis when no force, vertical force and slant force applied on liquid level.

Through the surface integration in the simulation, we can get the acoustic power of the spiral transducer at different axial distances from the upper liquid medium. We calculate the angular momentum of the acoustic vortex at different positions in the upper liquid according to the acoustic power of the acoustic beam. According to the simulation and calculation results, the variation of the acoustic power and angular momentum of the acoustic vortex with the axial distance can be obtained [Fig. S5]. When the axial distance increase, the angular momentum of the vortex is obviously smaller, which is due to the angular momentum transfer between the vortex and the liquid medium. It is precisely because the angular momentum transfer of the acoustic vortex, the velocity of the water layer increases and the pressure decreases greatly. So, compared with the non-spiral transducer, the single spiral-electrode transducer can lift the liquid column more easily.

In the experiment, we find that when the angle between the axis of the transducer and the interface of two-phase liquid is 45°, the liquid column can also be extracted at the interface of two-phase liquid. The liquid column is no longer vertical, but has a certain inclination, and the top of the liquid column always faces the transducer. As shown in Fig. 4(a), we show the comparison results of the working power of extracting the liquid column and droplets when transducers working vertically and obliquely. It is obvious that when the transducer is tilted, the liquid column and droplets can be extracted with less working power.

In this regard, we did a simple force analysis. When the acoustic radiation force first started to apply on the horizontal liquid interface, the surface tension produced on both sides of the point where transducer put radiation force on, and the radiation force is in the horizontal direction [Fig. 4(b)]. Due to the surface tension has no component in the vertical direction, it will not hinder the production of the liquid column.

When the liquid column is produced, the liquid interface shows raised state. Because the surface tension is always along the direction of liquid interface [Fig. 4(c)], it will produce downward surface tension along both sides of the liquid column. There is a vertical resultant force of the surface tension which is opposite to the direction of the radiation force, it will hinder the extraction of the liquid column.

When the slant transducer put radiation force on the liquid interface, the liquid column is slant because the radiation force will produce not only the vertical force, but also the horizontal force. It can be seen that the slant liquid column is smoother and the angle in the horizontal direction is smaller. It is more conducive that extract the liquid column, because the vertical resultant force of the surface tension is obviously reduced [Fig. 4(d)].

Here, we report a new kind of miniwatt aspirators which can accurately capture and manipulate wavelength-sized droplets. Inspired by the special spiral dynamic mechanism, we apply the acoustic vortex with OAM to the precise droplet capture and control in two-phase liquid for the first time. Through experiments, we use the acoustic vortex generated by single spiral-electrode transducer to extract droplets from the interface of two-phase liquids. Compared with the focused beam, the acoustic vortex can realize the extraction of wave droplets with less acoustic radiation force.

## 4. Conclusion

At present, the liquid-liquid two-phase fluid has a great advantage in droplet extraction because of its low impedance and small loss of acoustic waves when it penetrates the interface. However, there is no related research on the contactless extraction of droplets at the interface of two-phase fluids with higher impedance, such as liquid-gas. Due to the increase of the impedance of the two-phase interface, the droplet aspirator based on acoustic vortex is expected to break through the shackles of the gas-liquid interface to extract droplets. Miniwatt droplet aspirator has very important application value in chemical industry, medicine and other fields [47]. This special spiral structure is expected to be used in bio-inspired piezoelectric materials.

## APPENDIX A: MANUFACTURE OF SPIRAL PIEZOELECTRIC TRANDUCER

Our piezoelectric transducers are made from 25-mm radius and 1-mm thick PZT-4. We gave the thread model file drawn by COMSOL to the factory for electrode coating, with one side as spiral electrode and the other as fully coated electrode. After the electrode coating is completed, we need weld wires on both sides of the electrode as the input and output terminals of the transducers. Generally, the input terminal (red wire) is the spiral electrode and the output end (black wire) is the fully coated electrode. The thread of single spiral piezoelectric transducer can be directly welded to the wire. Since there are multiple threads in the multi-spiral piezoelectric transducer, it is necessary to first connect the threads through the wires in series and then weld the input terminal. Finally, the transducer that completes port welding can be made by gluing.

## APPENDIX B: THE EXPERIMENTAL DEVICES

In the experiments, we provided RF AC signals through Agelent 32210A function signal generator, and then used MODEL 2100L RF power amplifier with maximum voltage gain of 50 dB and frequency range of 10 kHz ~ 12 MHz for power amplification. Tektronix TDS2024B oscilloscope was used to measure the voltage at both ends of the transducer. The NI board used to measure the high-frequency current in the circuit is NI ELVIS II+.

## APPENDIX C: THE USE OF HIGH-SPEED CAMERA

We used Photron's FASTCAM SA1.1 high-speed camera to record the experiments. The camera is connected to a computer with a gigabit-nic through a cable, and the computer's IP address needs to be set correctly according to the camera manual. It is necessary to install the camera software PFV on the computer to control the camera for photographing and recording. Since we are used a black and white lens, we need to place a light source aimed at the camera lens to get the picture. In our experiments, the shooting frame rate of high-speed video was 5400 fps and the resolution was 1024×1024.

**Declaration of competing interest**

None.

**Acknowledgements**

We acknowledge Professor Jianchun Wang from Southern University of Science and Technology for his valuable suggestions on the part of our experimental analysis about Bernoulli's theorem. And we thank Robert Lirette and Likun Zhang for discussions by email. Funding: We also acknowledge support from the National Natural Science Foundation of China (Grant Nos.11772349 and 11972354). Author contributions: H.Z. conceived the idea.

H.Z., J.Y., G.Z. and Y.L. built up the whole system. H.Z. and J.Z. designed all the experiments. H.Z., Y.Z. and Y.C. carried out the corresponding numerical simulations. Y.Z. and B.B. experimented and filmed all the scenes with the high-speed camera for the shooting. H.Z. and Y.Z. contributed to the scientific presentation. All the authors approved the final version of the manuscript. Competing interests: All other authors declare that they have no competing interests. Data and materials availability: All data needed to evaluate the conclusions in the paper are present in the paper and/or the Supplementary Materials. Additional data related to this paper may be requested from the authors.